\documentstyle[12pt,epsfig]{article}
\def\lapprox{\lower .7ex\hbox{$\;\stackrel{\textstyle <}{\sim}\;$}}
\def\d{{\rm d}}

\def\smx{\stackrel{x\to 0}{\longrightarrow}}
\begin{document}
\begin{titlepage}
\vspace*{-1cm}
\begin{flushright}
DTP/95/62   \\
July 1995 \\
\end{flushright}
\vskip 1.cm
\begin{center}
{\Large\bf  Analytic Approaches to the Evolution of Polarised  \\[2mm]
Parton Distributions at Small {\it x}  }
\vskip 1.cm
{\large  T.~Gehrmann and W.J.~Stirling}
\vskip .4cm
{\it Departments of Physics and Mathematical Sciences, University of Durham \\
Durham DH1 3LE, England }\\
\vskip 1cm
\end{center}
\begin{abstract}
The $Q^2$ evolution of polarised parton distributions at small $x$ is studied.
Various analytic approximations are critically discussed. We compare
the full evolution with that obtained from the leading-pole approximation
to the splitting functions, and show that the validity of this approximation
depends critically on the $x \to  0$ behaviour of the starting distributions.
A new analytic solution which is valid at small $x$ is obtained, and its
domain of applicability is discussed.
\end{abstract}
\vfill
\end{titlepage}
\newpage

\section{Introduction}
The first moment of the polarised structure function $g_1(x,Q^2)$, the
Ellis-Jaffe sum rule \cite{ej}, determines the overall spin content of the
nucleon. Measurements of the Ellis-Jaffe  sum rule \cite{exp,slac}  involve an
extrapolation of $g_1$ for $x\rightarrow 0$, which is usually
performed by fitting a Regge-motivated form
\begin{equation}
g_1(x) = C x^{\alpha}
\label{one}
\end{equation}
to the experimental data points with the lowest $x$-values.
This procedure can be problematic, since these data points are
usually taken at a relatively low scale of $Q^2\sim 1 \;
\mbox{GeV}^2$, whereas the overall sum rule is evaluated at the average $Q^2$
of the experiment, which is typically between $3$ and $10\;\mbox{GeV}^2$.

In the recent past, various authors have attempted to calculate the
asymptotic behaviour of $g_1(x)$ (see for example Ref.~\cite{CR} for a
review of the various
 approaches). At scales of low momentum transfer
($Q^2 \approx 1 \; \mbox{GeV}^2$), a non--perturbative calculation
\cite{bass} of the flavour singlet contribution to $g_1$ shows good
agreement with $g_1^p$ at small $x$, but it should be noted that the
normalisation of this non--perturbative contribution is highly
sensitive to the only approximately known value of the vacuum quark
condensate.
The experimental discrepancy between $g_1^p$ and $g_1^d$ in
the small-$x$
region seems to contradict the above result. As the singlet distribution is
identical for both targets, this discrepancy indicates a sizeable
valence-quark contribution in this region.

With increasing $Q^2$, perturbative corrections  become more and
more important. These corrections affect both the valence and the
singlet contributions ($\Delta\Sigma = \sum_q(\Delta q + \Delta \bar q)$)
 to $g_1$ and give rise to  an evolution of the
corresponding parton densities \cite{ap}
\begin{eqnarray}
\frac{\partial}{\partial \ln Q^2} \Delta q_{val} (x,Q^2) & = &
\int_x^1 \frac{\d y}{y}  \Delta P_{qq} (y) \;\Delta q_{val} (x/y,Q^2)
\nonumber\\
\label{evo}
\frac{\partial}{\partial \ln Q^2}
\left(\begin{array}{c} \Delta \Sigma \\ \Delta G
\end{array} \right) (x,Q^2) & = &
\frac{\alpha_s(Q^2)}{2\pi}
\int_x^1 \frac{\d y}{y} \left(\begin{array}{cc} \Delta P_{qq}  & \Delta
P_{qg} \\ \Delta P_{gq} & \Delta P_{gg}\end{array} \right) (y)\;
\left(\begin{array}{c} \Delta
\Sigma \\ \Delta G \end{array} \right)(x/y,Q^2),\nonumber\\
\end{eqnarray}
without determining the densities themselves. These enter the above
equation in the form of the initial distributions
$\Delta q_{val}(x,Q_0^2)$, $\Delta \Sigma(x,Q_0^2)$ and $\Delta
G(x,Q_0^2)$, which form the boundary conditions for the solution.

In experimental measurements, these perturbative corrections are
incorporated by rescaling the value of $g_1$ to the average $Q^2$ of
the experiment. This rescaling procedure relies on the assumption that
the asymmetry $g_1(x) /F_1(x)$ satisfies exact Bjorken scaling, i.e. that the
$Q^2$-dependence of $g_1$ coincides with that   of $F_1$.
Although this assumption is consistent with the present data (which
cover only a small range of  $Q^2$ values  at fixed $x$), there is no
theoretical justification for it. In particular, examination of  the
polarised and unpolarised splitting functions \cite{ap} shows  that
$g_1(x) /F_1(x)$ should indeed show only a very weak $Q^2$ dependence
in the large-$x$ region, where both structure functions are dominated
by the valence quark content, as $\Delta P_{qq}(x)$ and $P_{qq}(x)$ are
identical. In contrast to this, the splitting functions in the singlet
sector, which dominates the small-$x$ behaviour of $F_1$, are different. The
unpolarised $P_{gq}(x)$ and $P_{gg}(x)$ have a soft gluon singularity at
$x=0$, which causes the steep rise of $F_1$ in the small-$x$
region. As this singularity is absent in the polarised splitting
functions (soft gluon emission does not  change the spin of the parent
parton), one would expect the ratio $\mid \!g_1(x) /F_1(x) \!\mid$ to
decrease with increasing $Q^2$.

With the exact splitting functions
\begin{eqnarray}
\Delta P_{qq}^{(f)} (x) & = & \frac{4}{3} \; \left[ 2\;
\frac{1}{(1-x)_{+}} - 1 -x + \frac{3}{2} \delta(1-x) \right] \nonumber \\
\Delta P_{qg}^{(f)} (x) & = & 2n_f \; \frac{1}{2} \; \left[ 2x-1 \right]
\nonumber\\
\Delta P_{gq}^{(f)} (x) & = & \frac{4}{3} \; \left[ 2-x \right]
\nonumber\\
\label{evoful}
\Delta P_{gg}^{(f)} (x) & = & 3 \; \left[  2\;
\frac{1}{(1-x)_{+}} + 2 -4x + \frac{11}{6} \delta(1-x)\right]
-\;\frac{n_f}{3} \delta(1-x)
\end{eqnarray}
it is not possible to find an analytic solution to (\ref{evo})
with realistic boundary conditions for the whole range of
$x$. By restricting themselves to small values of $x$ (although it is not
{\it a priori} clear which values of $x$ can be regarded as
 small), various
authors  have attempted to determine the asymptotic behaviour of $g_1$ in
the limit $Q^2\to \infty$. One possible approach \cite{CR} is to
assume that all the $Q^2$ dependence is dominated by the evolution of the
gluon, i.e. by $\Delta P_{gg}(x)$. This method gives successful
predictions for the unpolarised structure functions, due to the
$1/x$ pole in the unpolarised $P_{gg}$. As this pole is not present in
$\Delta P_{gg}$, the validity of this approach needs to  be
examined more carefully.

Another possible approach \cite{bf} to the asymptotic
small-$x$ behaviour is to transform (\ref{evoful}) into moment space
and to expand  around the rightmost singularity at $N=0$:
\begin{equation}
\langle \Delta P\rangle_N = {A\over N} + B + O(N) \Rightarrow
\Delta P(x) \approx A + B \delta(1-x) .
\end{equation}
This procedure  yields
the following approximate splitting functions\footnote{Similar splitting
functions containing only the residue at  $N=0$ were studied in
\protect{\cite{ber}}, giving qualitatively comparable
results to \protect{\cite{bf}}}:
\begin{eqnarray}
\Delta P_{qq}^{(l)} (x) & = & \frac{4}{3} \; \left[ 1 +\frac{1}{2}
\delta (1-x) \right] \nonumber \\
\Delta P_{qg}^{(l)} (x) & = & 2n_f \; \frac{1}{2} \; \left[ -1 + 2
\delta (1-x) \right]
\nonumber\\
\Delta P_{gq}^{(l)} (x) & = & \frac{4}{3} \; \left[ 2-\delta (1-x)
\right]
\nonumber\\
\label{bfevo}
\Delta P_{gg}^{(l)} (x) & = & 3 \; \left[  4 - \frac{13}{6}
\delta(1-x)\right]
-\;\frac{n_f}{3} \delta(1-x)
\end{eqnarray}
With these simplified splitting functions, one can  analytically
solve (\ref{evo}) for asymptotic values of $Q^2$ with realistic
boundary conditions in the small-$x$ region. This approach is based on the
fact that the behaviour of the parton distributions at small $x$ is
governed by the  region around $N=0$ in moment space. This property
can be understood from the $N$-singularity structure of the initial
distributions: a logarithmic ($\sim 1/x$) singularity
 coincides with a pole at $N=0$ in the moment transform,
 a power-like singularity of the
 form $x^{\alpha}$ transforms
into $\Gamma(\alpha+N)$, which has a singularity at $N = -\alpha$ (see
Fig.~1). It is important to notice, however, that the expansion around the
$N=0$ pole in moment space agrees with the full splitting function
only within a circle of unit radius (Fig.~1).
Outside this circle, the series might
still be convergent, but its value will be different from that given by
 the full splitting
function. This especially affects the reliability of this approach for
low values of $\alpha$.
In the extreme case $\alpha$ could approach $-1$
giving rise to  a pole close to the boundary of the circle of convergence.

In this letter, we examine the validity of analytical approaches to
the small-$x$ behaviour of $g_1$. Section~2 contains a
study of the evolution matrix on the right-hand side
of (\ref{evo}). Its properties in
the case of power-like  ($x^{\alpha}$) boundary conditions are discussed,
 using the full
and the leading-pole expanded splitting functions. By examining the
sensitivity of the evolution matrix to the form of the parton
distributions in the large-$x$ region, we are able to assess
 when $x$ can be
regarded as small. In Section~3 we present an analytic solution of
(\ref{evo}), which becomes exact for $x\to 0$. Finally, Section~4
contains some phenomenological implications and conclusions.

\section{Study of the evolution matrix}

Several qualitative features of the polarised parton densities can
already be determined by inserting simple test distributions
in the right-hand side of Eq.~(\ref{evo}). The resulting elements of
the evolution matrix determine the local change of the parton
densities with increasing $\ln (Q^2)$. Furthermore, for
$Q^2/Q_0^2$ not too large one can approximate
the solution of the Altarelli-Parisi equations by
\begin{eqnarray}
\left(\begin{array}{c} \Delta \Sigma \\ \Delta G
\end{array} \right) (x,Q^2) & = & \left(\begin{array}{c} \Delta \Sigma  \\
\Delta G \end{array} \right) (x,Q_0^2)
\nonumber\\
& + &  \frac{\alpha_s(Q_0^2)}{2\pi}
\int_x^1 \frac{\d y}{y}
\left(\begin{array}{cc} \Delta P_{qq}  & \Delta
P_{qg} \\ \Delta P_{gq} & \Delta P_{gg}\end{array} \right)
(y)  \left(\begin{array}{c} \Delta
\Sigma \\ \Delta G \end{array} \right)(x/y,Q_0^2) \; \ln
\left(\frac{Q^2}{Q_0^2}\right).
\end{eqnarray}

A realistic choice of test distribution is
\begin{equation}
t(x) = x^{\alpha} (1-x)^{\beta}\qquad \mbox{with} \quad
(-1<\alpha<0,\;\beta>0),
\end{equation}
which is similar to the analytic forms of the parton densities at $Q_0^2$
 used in
recent fits to the polarised structure function data \cite{bf,gs}. The
exponent $\alpha$ determines the behaviour of the distribution in the
small-$x$ regime, whereas the large-$x$ behaviour is controlled by
$\beta$.
 Variations of $\beta$ should therefore not affect any
predictions of the small-$x$ behaviour of the parton distributions.
This property can be used to define the range of validity of these
predictions, i.e. to indicate if $x$ can be regarded as small or not.

The elements of the evolution matrix
\begin{equation}
A_{ij} = \int_x^1 \frac{\d y}{y}\; \Delta P_{ij} (y) \; t\left(\frac{x}{y}
\right)
\end{equation}
can be computed analytically. The necessary integrals are given in
Appendix~A.
Using the full splitting functions (\ref{evoful}), we
find\footnote{For simplicity, we take $n_f=3$ throughout this study}
\begin{eqnarray}
A_{qq}^{(f)} (x) & = & \frac{4}{3} \left[ 2 A_1(x) -A_2(x) -A_3(x) +
\frac{3}{2} A_4(x) \right] \nonumber \\
A_{qg}^{(f)} (x) & = & 3 \left[-A_2(x) + 2 A_3(x) \right]
\nonumber \\
A_{gq}^{(f)} (x) & = & \frac{4}{3} \left[ 2 A_2(x) -A_3(x) \right]
\nonumber \\
A_{gg}^{(f)} (x) & = & 3 \left[ 2 A_1(x) + 2 A_2(x) - 4 A_3(x) +
\frac{3}{2} A_4(x) \right],
\end{eqnarray}
while the leading-pole expanded \cite{bf} splitting functions
of (\ref{bfevo}) yield
\begin{eqnarray}
A_{qq}^{(l)} (x) & = & \frac{4}{3} \left[ A_2(x) + \frac{1}{2} A_4(x)
\right] \nonumber\\
A_{qg}^{(l)} (x) & = & 3 \left[ -A_2(x) + 2 A_4(x)
\right] \nonumber\\
A_{gq}^{(l)} (x) & = & \frac{4}{3} \left[ 2 A_2(x) -A_4(x)
\right] \nonumber\\
A_{gg}^{(l)} (x) & = & 3 \left[ 4 A_2(x) - \frac{5}{2} A_4(x) \right].
\end{eqnarray}

A closer inspection of the $A_{ij}$ shows that all of them diverge
like $x^{\alpha}$ as $x\to 0$. The behaviour in the limit $x \to 0$ can
therefore be written as
\begin{equation}
\label{limit}
\lim_{x \to 0} A_{ij} (x) = a_{ij} x^{\alpha}.
\end{equation}
Provided that both the initial quark singlet and the initial gluon
distributions have power-like boundary conditions in the limit $x\to 0$,
these most singular terms will dominate the right-hand side
of (\ref{evo}). The replacement of the $A_{ij}^{(f)}$ by the above
expressions (\ref{limit}) in (\ref{evo}) should therefore enable us to find an
analytic solution for $\Delta \Sigma(x,Q^2)$ and $\Delta G(x,Q^2)$,
which becomes exact for  $x\to 0$. This exercise will be performed in
the following section.

The $a_{ij}$ coefficients for the full and the leading-pole expanded
splitting functions are {\it not} identical:
\begin{equation}
\begin{array}{ll}
\displaystyle
a_{qq}^{(f)}  =  \frac{4}{3} \;\left[2(-\psi(-\alpha) - \gamma_E) +
\frac{1-2\alpha}{\alpha(1-\alpha)}+ \frac{3}{2} \right] \qquad &
\displaystyle
a_{qq}^{(l)}  =  \frac{4}{3} \; \frac{-2+\alpha}{2\alpha}
\\
\vspace{3mm}
\displaystyle
a_{qg}^{(f)}  =  3 \; \frac{1+\alpha}{\alpha(1-\alpha)} \qquad &
\displaystyle
a_{qg}^{(l)}  =  3 \; \frac{1+2\alpha}{\alpha}
\\
\vspace{3mm}
\displaystyle
a_{gq}^{(f)}  =  \frac{4}{3} \;\frac{-2+\alpha}{\alpha(1-\alpha)} \qquad &
\displaystyle
a_{gq}^{(l)}  =  \frac{4}{3} \; \frac{-2-\alpha}{\alpha}
\\
\vspace{3mm}
\displaystyle
a_{gg}^{(f)}  = 3 \; \left[ 2(-\psi(-\alpha) - \gamma_E)
- \frac{2+2\alpha}{\alpha(1-\alpha)} + \frac{3}{2} \right] \qquad &
\displaystyle
a_{gg}^{(l)}  = 3 \; \frac{-8-5\alpha}{2\alpha}
\end{array}
\end{equation}
Here $\psi(x)$ is the usual psi (digamma) function \cite{AbSteg}.

Figure~2 shows the $A_{ij}^{(f)}$ for $\alpha=-0.25,-0.6$ and $\beta=4,9$,
together with the approximate forms $A_{ij}^{(l)}$ and the limits
$a_{ij}^{(f)}x^{\alpha}$. This figure displays the following important
features of the evolution matrix in the small-$x$ region:
\begin{itemize}
\item[{(i)}] Although the test distributions $x^{\alpha}(1-x)^4$ and
$x^{\alpha}(1-x)^9$ differ by less than 5\% for $x\le 0.01$, the
corresponding $A_{ij}^{(f)}$ differ by up to a factor of 2 in the same
range. This clearly demonstrates that even at $x=0.01$ and below the
evolution is sensitive to the behaviour of the parton distributions in
the large-$x$ region.
The insensitivity of the $A_{ij}^{(f)}$
to variations of $\beta$ can furthermore be used to define whether
$x$ can be regarded as small. For example,
by requiring $A_{ij}^{(f)}$ to vary by
less than 30\% for all combinations in $i$ and $j$ and both values of
$\alpha$, we find that only $x\le 0.001$ can be regarded as small, and the
more conservative bound of less than 10\% deviation yields $x \le
0.0001$. It should therefore be clear that the mere knowledge of
$g_1$ at the lowest $x$ values accessible with fixed-target experiments
is insufficient to predict the asymptotic behaviour of $g_1$ in the
small-$x$ limit, as the behaviour of the parton distributions at these
values of $x$ is still closely correlated with the distributions in
the large-$x$ region.

\item[{(ii)}] The convergence of the $A_{ij}^{(f)}$ towards
$a_{ij}^{(f)}x^{\alpha}$ improves for smaller values of $\alpha$.
This behaviour just reflects the fact that
$A_{ij}^{(f)}$ contains, in
addition to this leading term, less singular terms proportional to $\ln(x)$.
 In general, these  lower $\mid \! A_{ij}^{(f)} \! \mid$.
If $t(x)$ is less singular that $x^{-1/e}$, the logarithmic terms
are larger than the power-like terms for
\begin{equation}
x > x_0 (\alpha) = \left( \frac{\omega(\alpha)}{\alpha}
\right)^{\frac{1}{\alpha}}
\end{equation}
where $\omega(\alpha)$ is the branch of Lambert's $\omega$-function
which satisfies $\omega(-1/e)=-1$.
 As $x_0$ decreases very quickly
with $\alpha$
($x_0 \approx 10^{-15}$ for $\alpha=-0.1$), the replacement
$A_{ij}^{(f)}(x) \to a_{ij} x^{\alpha}$, although formally still
correct, loses its meaning for values of $\alpha$ close to 0 in any
physically relevant region.

\item[{(iii)}] While the $A_{ij}^{(l)}$ resemble the $A_{ij}^{(f)}$
for values of  $\alpha$ close to 0, they disagree for smaller
$\alpha$. This feature becomes most striking for the $A_{qg}$ (see Fig.~2).
The full splitting functions \cite{ap} predict that a
positive gluon polarisation in the small-$x$ region will always
generate a negative contribution to the sea polarisation. In contrast,
  the leading-pole expanded
splitting functions of \cite{bf} predict a {\it positive} sea polarisation,
if the gluon polarisation $\Delta G(x)$ is more singular than
$x^{-0.5}$.
This behaviour can be inferred from the $\alpha$
dependence of the $a_{ij}$ displayed in Fig.~3.
The good agreement
for higher values of $\alpha$ is due to the fact that all leading
contributions in $\ln (x)$ are contained in the $N=0$ pole and hence
are well approximated by the $A_{ij}^{(l)}$. As elaborated above, these
contributions remain important for a finite range in $x > x_0 >0$.
The asymptotic predictions of \cite{bf} will therefore still
approximate the full evolution, provided they are restricted to this
finite range.

\item[{(iv)}] The magnitude of $A_{gg}$ is larger by a factor 3
 than the magnitude of all the other terms, but $A_{gg}$ is not more
singular than any other contribution. Therefore, the small-$x$
estimate of Ref.~\cite{CR}  is quantitative at best, and should be
expected to yield a less accurate prediction than the corresponding
estimate of the unpolarised distributions.
\begin{table}[htb]
\begin{center}
\begin{tabular}{|c|c|c|c|} \hline
\rule[-1.2ex]{0mm}{4ex}& $N\le -2$ & $N=-1$ & $N=0$ \\ \hline
\rule[-1.2ex]{0mm}{4ex}$3/4 \;\Delta P_{qq}$ & 2 & 1 & 1 \\ \hline
\rule[-1.2ex]{0mm}{4ex}$1/3 \;\Delta P_{qg}$ & 0 & 2 & -1 \\ \hline
\rule[-1.2ex]{0mm}{4ex}$3/4 \;\Delta P_{gq}$ & 0 & -1 & 2 \\ \hline
\rule[-1.2ex]{0mm}{4ex}$1/3 \;\Delta P_{gg}$ & 2 & -2 & 4 \\ \hline
\end{tabular}
\caption{Residues of the polarised splitting functions in $N$-moment
space. The residues for all negative integers with $N\le -2$ are
identical.}
\label{tab:res}
\end{center}
\end{table}
\item[{(v)}] The agreement between leading pole expanded and full
splitting
functions is better for the $A_{gq}$ and $A_{gg}$ than it is for
$A_{qq}$ and $A_{qg}$. This feature can be understood from the
relative magnitude of the residues in the corresponding splitting
functions (Table \ref{tab:res}): the $N=0$ residue is dominant only in the
$P_{gq}$ and $P_{gg}$ splitting functions, the other two splitting
functions contain residues for $N<0$, which are twice as big as the
$N=0$ residue.
\end{itemize}

It should be clear from the above that the leading-pole expansion
of Ref.~\cite{bf} gives a reliable approximation to the evolution matrix
in the small-$x$ region,
provided that the initial distributions are significantly
less singular than $x^{1/e}$. For more singular distributions, this
approach results in a manifestly  different evolution matrix and
hence will yield a different small-$x$ behaviour of the polarised parton
distributions.

\section{Solution of the Altarelli-Parisi equations in the limit $x\to 0$}

Provided both polarised singlet quark and gluon densities have
power-like boundary
conditions in the small-$x$ region,
\begin{equation}
\Delta \Sigma (x,Q_0^2) \sim x^{\alpha_q} , \quad \Delta G (x,Q_0^2) \sim
x^{\alpha_G} \qquad \mbox{with } -1 < \alpha_q,\alpha_G <0 ,
\end{equation}
one can find a solution of the Altarelli-Parisi equations which
becomes exact in the limit $x\rightarrow 0$ and has the form
\begin{eqnarray}
\Delta q_{val} (x,Q^2) & = & R_v(Q^2,Q_0^2)\; x^{\alpha_v}\nonumber\\
\Delta \Sigma (x,Q^2) &=& R_{qq}(Q^2,Q_0^2) x^{\alpha_q} + R_{qg}(Q^2,Q_0^2)
x^{\alpha_G}, \nonumber\\
\Delta G (x,Q^2) &=& R_{gq}(Q^2,Q_0^2) x^{\alpha_q} + R_{gg}(Q^2,Q_0^2)
x^{\alpha_G}.
\end{eqnarray}
A detailed derivation of this solution and the explicit forms of the
$R$-functions is given in  Appendix~B.

The above bounds on $\alpha$ cover the whole theoretically
allowed range: as the first moments of the distributions have to be
finite, we find $\alpha>-1$. Furthermore, inspection of the singularity
structure of the evolution equations (Fig.~1) shows that any initial
distribution, which is finite in the small-$x$ region, will develop a
logarithmic divergence due to the $N=0$ singularity of the splitting
functions. The case of finite or logarithmic boundary conditions
can be treated correctly with the leading-pole approximation -- its
asymptotics are discussed in \cite{bf}.
In a previous analysis \cite{gs} of the experimental data on polarised
structure
functions we have found $\alpha_q=\alpha_v\simeq -0.55$.
The experimental data used in this analysis were insufficient to determine
$\alpha_G$, and therefore it was fixed to be $0$. In contrast to this,
more recent measurements at lower $Q^2$ \cite{slac} favour $\alpha_G<0$.

As we have neglected all contributions of order $\ln (x)$ in the
above solution, we expect it  to be reliable only for $x <
x_0(\mbox{max}(\alpha_q, \alpha_G))$. In order to compare this
approach with the leading pole expansion of Ref.~\cite{bf} and the
numerical solution of (\ref{evo}) with the full splitting functions,
we have evaluated the distributions
 for $Q_0^2=4\;\mbox{GeV}^2$ and
$Q^2=100\;\mbox{GeV}^2$,  using $n_f=3, \Lambda^{QCD}=200
\;\mbox{MeV}$ and  the following initial distributions:
\begin{eqnarray}
\Delta \Sigma(x,Q_0^2)  & = & N_q x^{\alpha_q} (1-x)^\beta \nonumber \\
\Delta G(x,Q_0^2)  & = & N_G x^{\alpha_G} (1-x)^\beta  \\
\Delta q_{val}(x,Q_0^2)  & = & N_{val} x^{\alpha_v} (1-x)^\beta .\nonumber
\end{eqnarray}
To illustrate the validity of the various approximations, we adopt the
following parameter values: $\alpha_q,\alpha_G,\alpha_v = -0.6, -0.25$,
$\beta=4,9$, and for simplicity we take $N_q = N_g = N_v = 1$.

Figures~4 (a), (b) and  (c) show the behaviour of the gluon, singlet quark and
valence quark distributions respectively, at small $x$ and $Q^2 = 4,\;
100$~GeV$^2$. The initial distributions $x^{\alpha} (1-x)^4$ are
indicated as solid lines.

Starting with the gluon distribution (Fig.~4(a)),
we see that for $ x < 10^{-2}$, the leading-pole approximation
to the splitting funcitons (dotted lines) gives
excellent agreement with the full evolution (dashed line), especially
for values  of $\alpha_q, \alpha_G$ close to 0. This is consistent
with the agreement between the corresponding $A_{gg}$ functions
shown in Fig.~2 and can be understood from the $N=0$ dominance in the
$\Delta P_{gq}$ and $\Delta P_{gg}$ splitting functions.
 In contrast, the $x^\alpha$
approximation (short-dashed line) significantly overestimates the
evolution in the $x$ range shown, espcially for
$\alpha_q, \alpha_G$ close to 0. Convergence of this approach can only
be observed at even smaller values of $x$.
Note, however,  the sensitivity to the large-$x$
behaviour. While both the dotted and the dashed lines are computed
with $\beta = 4$, the dot-dashed curve
corresponds to full splitting
function evolution for $\beta = 9$, i.e. a softer large-$x$
distribution.  Evidently there is a significant sensitivity to the
behaviour at large $x$ even for $x$ values as small as $O(10^{-3})$.
This casts doubt on the idea of using data on the evolution of the
small-$x$ structure functions alone to determine the gluon
distribution.

For the singlet quark distribution (Fig.~4(b)) the situation is
rather different. Here the leading-pole approximation {\it
overestimates} the evolution at small $x$. This is readily understood
from the behaviour of the corresponding $A_{qq}$ and $A_{qg}$ functions
in Fig.~2, both of which are systematically more positive for
the leading-pole splitting functions. In fact we see that for
$\alpha_q = -0.25$ and $\alpha_G = -0.6$, the full evolution gives a
negative singlet distribution at small  $x$, whereas the leading pole
splitting functions give a positive distribution. Notice also that
the evolution is less sensitive to the large-$x$ behaviour (compare
the dashed and dot-dashed curves which correspond to $\beta = 4,9$
respectively) than for the gluon distribution.  For $\alpha_q = \alpha_G
= -0.6$, the $x^\alpha$ approximation is quite reasonable, and certainly
better than the leading-pole approximation. However the opposite is true
when both $\alpha_q, \alpha_G$ are close to $0$.

Finally, Fig.~4(c) compares the valence quark evolution in the various
approximations. This depends only on $\Delta P_{qq}$, and so the
behaviour here is a direct reflection of the corresponding $A_{qq}$
shown in Fig.~2.  In particular, for $\alpha_q = -0.6$ the $x^\alpha$
approximation is very good, while
the leading-pole approximation overestimates  the evolution  at all
$x$ values shown.
For less singular small-$x$ behaviour ($\alpha_q = -0.25$), however,
both approximations reproduce the full evolution, the leading-pole
approximation showing slightly better convergence for $x> O(10^{-4})$.

In practice, the normalisations of the singlet quark and gluon
distribution, $N_q$ and $N_G$, will not be the same.
As the evolution of the gluon densitity is dominated by the
gluon-to-gluon splitting, it will be almost unaffected by changes of
$N_q$. Only if $N_q$ is one or more orders of magnitude larger than
$N_G$, will the impact of quark-to-gluon splitting become visible.
More drastic effects of a change in the relative normalisation can be
expected for the quark singlet distribution, as contributions from
quark-to-quark and gluon-to-quark splitting have the same magnitude
but opposite signs (cf.~Fig.~3). Therefore, a relative increase of
$N_G$ yields a faster evolution of the quark distribution to negative values.

The
convergence properties of the different analytic approaches are
almost unaffected by changes in the normalisation. Only for $N_G \gg
N_q$ do  we find that convergence of the $x^{\alpha}$ approximation to
the singlet distribution sets in for smaller values of $x$.
This simply reflects an increased impact of the gluon-to-gluon
splitting.

\section{Conclusions}
In this letter we have studied the feasibility of two different analytic
approaches to the evolution of polarised parton densities at small
$x$, finding that none of these approaches is able to give reliable
predictions for the whole theoretically allowed range of boundary
conditions in the small-$x$ region. In the leading-pole expansion
\cite{bf,ber}, the full splitting
functions $\Delta P_{ij}$ are replaced by the leading terms of their
Laurent series around $N=0$. As this approach correctly reproduces
 all
terms proprotional to $\ln x$ generated in the evolution, it is found
to be in good agreement with the full evolution if the initial quark
and gluon distributions are less singular than $x^{-1/e}$. For more
singular boundary conditions, only the gluon distribution is
reproduced correctly, in particular
the quark distribution is overestimated. Keeping
only terms with powerlike singularities in the evolution equation, we
were able to derive an exact solution of this equation in the limit $x
\to 0$. As we have neglected all logarithmic terms in this approach,
its convergence is best for boundary conditions of quark and gluon
distributions
more singular than $x^{-1/e}$. For less singular boundary conditions,
this approach still converges towards the full  solution, but its
predictions are far away from the full solution for any realistic
experimental value of $x$.

We have also  shown that
the evolution of the polarised gluon distribution is sensitive to the
shape of this distribution in the large-$x$ region. This observation
raises doubts on the possibility of determining the gluon polarisation
 from the
evolution of $g_1$ in the small-$x$ region. It furthermore demonstrates
the need for complementary measurements of $\Delta G(x)$ (e.g.
 from $J/\Psi$-production or direct-$\gamma$ measurements).

We have seen that the effects of the evolution on the quark
distributions in the small-$x$ region are rather small, as the quark-to-quark
and the gluon-to-quark splitting contribute with opposite signs. The
gluon distribution is indeed rising with increasing $Q^2$, but  only
contributes to $g_1$ at order $\alpha_s(Q^2)$.
Bearing in mind that $\Delta G$ contributes with a negative coefficient
function to $g_1$, one expects that $g_1$ will become negative at
small $x$ for asymptotic values of $Q^2$, due to the gluonic
contribution and the negative sea polarisation generated from
$g \to q \bar q$ splitting.

In general, the effects of the evolution on the polarised parton
densities will be more moderate than the effects on the unpolarised
densities. The assumption of approximate scaling for $g_1(x)/F_1(x)$ in
the small-$x$ region is therefore rather doubtful. It seems more
realistic to assume approximate scaling for $g_1(x)$ for the range of
fixed-target experiments,  due to the
partial cancellation of quark and gluon evolution as explained
above.

\section*{Acknowledgements}

\noindent  Financial support from  the UK PPARC (WJS), and from
the Gottlieb Daimler- und Karl Benz-Stiftung and the
Studienstiftung des deutschen Volkes (TG) is gratefully acknowledged.
This work was supported in part by the EU Programme
``Human Capital and Mobility'', Network ``Physics at High Energy
Colliders'', contract CHRX-CT93-0357 (DG 12 COMA).
\goodbreak

\begin{appendix}
\section{Convolution integrals of the test distribution}

For the test distribution
\begin{equation}
t(x) = x^{\alpha} (1-x)^{\beta}\qquad \mbox{with} \quad
(-1<\alpha<0,\;\beta>0),
\end{equation}
the convolution integrals
\begin{equation}
A_{ij} = \int_x^1 \frac{\d y}{y}\; \Delta P_{ij} (y) \; t\left(\frac{x}{y}
\right)
\end{equation}
on the right-hand side of (\ref{evo}) can be expressed in an
analytic form. From the explicit forms of the splitting functions
given in (\ref{evoful}) and (\ref{bfevo}), one sees that the
required integrals are
\begin{eqnarray}
A_1 (x) & = &  \int_x^1 \frac {\d y}{y}\; \frac{1}{(1-y)_{+}}\; \left(
\frac{x}{y} \right)^{\alpha} \; \left( 1- \frac{x}{y} \right)^{\beta}
\nonumber\\
& = & x^{\alpha} (1-x)^{\beta} \big[ \ln (1-x) + \frac{\alpha + \beta
+1}{ \beta + 1} (1-x) \;_{3}F_{2}
(2+\beta+\alpha,1,1;2,2+\beta;(1-x))\ \nonumber \\
& & -\psi(\beta+1) -\gamma_E\big]
\nonumber\\
A_2 (x) & = & \int_x^1 \frac {\d y}{y}\; \left(
\frac{x}{y} \right)^{\alpha} \; \left( 1- \frac{x}{y} \right)^{\beta}
\nonumber\\
& = & (1-x)^{\beta+1} \frac{1}{\beta+1} \;_{2}F_{1}
(1-\alpha,1+\beta;2+\beta;(1-x)) \nonumber\\
A_3 (x) & = & \int_x^1 \frac {\d y}{y}\; y \left(
\frac{x}{y} \right)^{\alpha} \; \left( 1- \frac{x}{y} \right)^{\beta}
\nonumber\\
& = & x (1-x)^{\beta+1} \left[\frac{1}{1-\alpha}
x^{\alpha-1} - \frac{\alpha+\beta}{(1-\alpha)(\beta+1)} \;_{2}F_{1}
(1-\alpha,1+\beta;2+\beta;(1-x))\right] \nonumber\\
A_4 (x) & = &  \int_x^1 \frac {\d y}{y}\; \delta (1-y) \left(
\frac{x}{y} \right)^{\alpha} \; \left( 1- \frac{x}{y} \right)^{\beta}
\nonumber\\
& = & x^{\alpha} (1-x)^{\beta}.
\end{eqnarray}
All these functions diverge like $x^{\alpha}$ as $x\to 0$,
and the leading singular behaviour at small $x$ is found to be
\begin{eqnarray}
A_1(x) & \smx & x^{\alpha} \left[ -\psi(-\alpha) - \gamma_E \right]
\nonumber\\
A_2(x) & \smx & \frac{1}{-\alpha} x^{\alpha} \nonumber\\
A_3(x) & \smx & \frac{1}{1-\alpha} x^{\alpha} \nonumber\\
A_4(x) & \smx & x^{\alpha}.
\end{eqnarray}

\section{Analytic solution of the Altarelli-Parisi equations for
$x\to 0$ }

A solution of the
Altarelli--Parisi evolution equations (\ref{evo}) can never be more
singular at $x=0$ than the starting distributions. It follows that the most
singular parts of the valence quark, singlet quark and gluon
distributions can be obtained by inserting the following ansatz
\begin{eqnarray}
\Delta q_{val} (x,Q^2) & = & R_v(Q^2,Q_0^2)\; x^{\alpha_v}\nonumber\\
\Delta \Sigma (x,Q^2) & = & R_{qq}(Q^2,Q_0^2) \;x^{\alpha_q} +
R_{qg}(Q^2,Q_0^2) \;
x^{\alpha_G}, \nonumber\\
\Delta G (x,Q^2) & = & R_{gq}(Q^2,Q_0^2) \;x^{\alpha_q} +
R_{gg}(Q^2,Q_0^2)\;
x^{\alpha_G}
\end{eqnarray}
into (\ref{evo}).
Keeping only terms proportional to $x^{\alpha_v}$, $x^{\alpha_q}$
and $x^{\alpha_G}$ on the right-hand side, we obtain the following
evolution equations for the $R$ coefficients ($\beta_0 = 11 - 2/3
n_f$) :
\begin{eqnarray}
\frac{\partial}{\partial \ln \alpha_s} R_v (Q^2,Q_0^2) & = & - \;
\frac{2}{\beta_0} a_{qq}(\alpha_v) R_v (Q^2,Q_0^2) \nonumber \\
 \frac{\partial}{\partial \ln \alpha_s}
\left(\begin{array}{c} R_{qq} \\ R_{gq}
\end{array} \right) (Q^2,Q_0^2) & = & - \; \frac{2}{\beta_0}
 \left(\begin{array}{cc} a_{qq} (\alpha_q)  & a_{qg}
(\alpha_q) \\
a_{gq} (\alpha_q)& a_{gg} (\alpha_q)
\end{array} \right)\;
\left(\begin{array}{c} R_{qq} \\ R_{gq} \end{array}
\right)(Q^2,Q_0^2)\nonumber \\
 \frac{\partial}{\partial \ln \alpha_s}
\left(\begin{array}{c} R_{qg} \\ R_{gg}
\end{array} \right) (Q^2,Q_0^2) & = & - \; \frac{2}{\beta_0}
 \left(\begin{array}{cc} a_{qq} (\alpha_G)  & a_{qg}
(\alpha_G) \\
a_{gq} (\alpha_G)& a_{gg} (\alpha_G)
\end{array} \right)\;
\left(\begin{array}{c} R_{qg} \\ R_{gg}\end{array}
\right)(Q^2,Q_0^2).
\end{eqnarray}
As we are interested in the asymptotic solution for the {\it full} splitting
functions,  all $a_{ij}$ in the above are $a_{ij}^{(f)}$.

Introducing
\begin{eqnarray}
s & = & \ln \left( \frac{\displaystyle \ln (Q^2/\Lambda^2)}{\displaystyle
\ln (Q_0^2/\Lambda^2)}\right) \nonumber\\
\omega_{\pm} (\alpha) & = & \frac{1}{2} \left( a_{qq} (\alpha) +
a_{gg} (\alpha) \pm \sqrt{ \left( a_{qq} (\alpha) -a_{gg}
(\alpha) \right)^2 + 4  a_{gq} (\alpha) a_{qg}
(\alpha)}\right),
\end{eqnarray}
the general solution of these equations reads
\begin{eqnarray}
R_v(Q^2,Q_0^2) & = & N_v  \;\exp \left\{ \frac{2}{\beta_0} a_{qq}
(\alpha_v) s \right\} \nonumber\\
R_{qq}(Q^2,Q_0^2) & = & R_{qq+} (Q_0^2) \;\exp \left\{ \frac{2}{\beta_0}
\omega_{+} (\alpha_q) s \right\}  +  R_{qq-} (Q_0^2) \;\exp \left\{
\frac{2}{\beta_0} \omega_{-} (\alpha_q) s \right\} \nonumber\\
R_{gq}(Q^2,Q_0^2) & = & R_{gq+} (Q_0^2) \;\exp \left\{ \frac{2}{\beta_0}
\omega_{+} (\alpha_q) s \right\}  +  R_{gq-} (Q_0^2) \;\exp \left\{
\frac{2}{\beta_0} \omega_{-} (\alpha_q) s \right\} \nonumber\\
R_{qg}(Q^2,Q_0^2) & = & R_{qg+} (Q_0^2) \;\exp \left\{ \frac{2}{\beta_0}
\omega_{+} (\alpha_G) s \right\}  +  R_{qg-} (Q_0^2) \;\exp \left\{
\frac{2}{\beta_0} \omega_{-} (\alpha_G) s \right\} \nonumber\\
R_{gg}(Q^2,Q_0^2) & = & R_{gg+} (Q_0^2) \;\exp \left\{ \frac{2}{\beta_0}
\omega_{+} (\alpha_G) s \right\}  +  R_{gg-} (Q_0^2) \;\exp \left\{
\frac{2}{\beta_0} \omega_{-} (\alpha_G) s \right\},
\end{eqnarray}
where the $R_{ij\pm} (Q_0^2)$ are determined by the boundary
conditions at $Q_0^2$. As we assume that
the initial distributions for the
quark singlet and the gluon have the form
\begin{equation}
\Delta \Sigma (x,Q_0^2) = N_q \; x^{\alpha_q} , \quad \Delta G
(x,Q_0^2) = N_G \;x^{\alpha_G},
\end{equation}
these constants are determined to be
\begin{equation}
\begin{array}{ll}
\displaystyle
R_{qq+} (Q_0^2) = \frac{\displaystyle \omega_{+} (\alpha_q) -
a_{gg}(\alpha_q)}{\displaystyle \omega_{+} (\alpha_q) - \omega_{-}
(\alpha_q)} N_{q}, & R_{qq-} (Q_0^2) = - \; \frac{\displaystyle
\omega_{-} (\alpha_q) - a_{gg}(\alpha_q)}
{\displaystyle \omega_{+} (\alpha_q) - \omega_{-}
(\alpha_q)} N_{q},\\
\vspace{3mm}
R_{gq+} (Q_0^2) = \frac{\displaystyle a_{gq} (\alpha_q)}
{\displaystyle \omega_{+} (\alpha_q) - \omega_{-}
(\alpha_q)} N_{q}, & R_{gq-} (Q_0^2) = - \; \frac{\displaystyle
a_{gq} (\alpha_q)}
{\displaystyle \omega_{+} (\alpha_q) - \omega_{-}
(\alpha_q)} N_{q},\\
\vspace{3mm}
R_{qg+} (Q_0^2) = \frac{\displaystyle a_{qg} (\alpha_G)}
{\displaystyle \omega_{+} (\alpha_G) - \omega_{-}
(\alpha_G)} N_{g}, & R_{qg-} (Q_0^2) = - \; \frac{\displaystyle
a_{qg} (\alpha_G)}
{\displaystyle \omega_{+} (\alpha_G) - \omega_{-}
(\alpha_G)} N_{g},\\
\vspace{3mm}
R_{gg+} (Q_0^2) = \frac{\displaystyle \omega_{+} (\alpha_G) -
a_{qq}(\alpha_G)}{\displaystyle \omega_{+} (\alpha_G) - \omega_{-}
(\alpha_G)} N_{g}, & R_{gg-} (Q_0^2) = - \; \frac{\displaystyle
\omega_{-} (\alpha_G) - a_{qq}(\alpha_G)}
{\displaystyle \omega_{+} (\alpha_G) - \omega_{-}
(\alpha_G)} N_{g}.
\end{array}
\end{equation}

\end{appendix}

\vfill
\newpage
\subsection*{Figure Captions}

\begin{description}

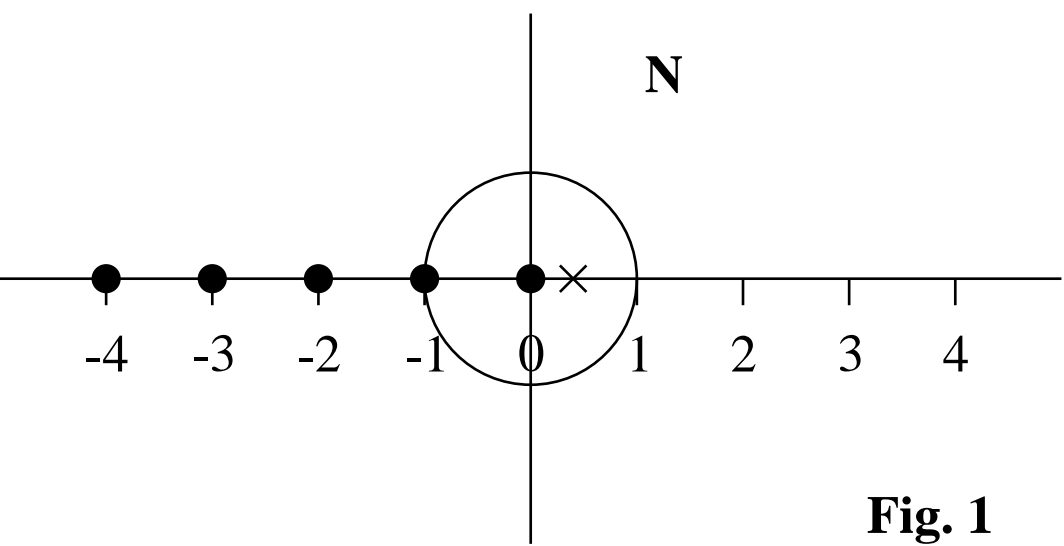
\item[Figure 1 ] Singularity structure of the evolution equations in
the complex $N$-moment plane. Dots ($\bullet$) denote the poles
of the splitting functions, and the cross ($\times$) indicates the
small-$x$ singularity of the initial distribution. The leading-pole
expansion only converges to the splitting function in the unit
circle around the origin.

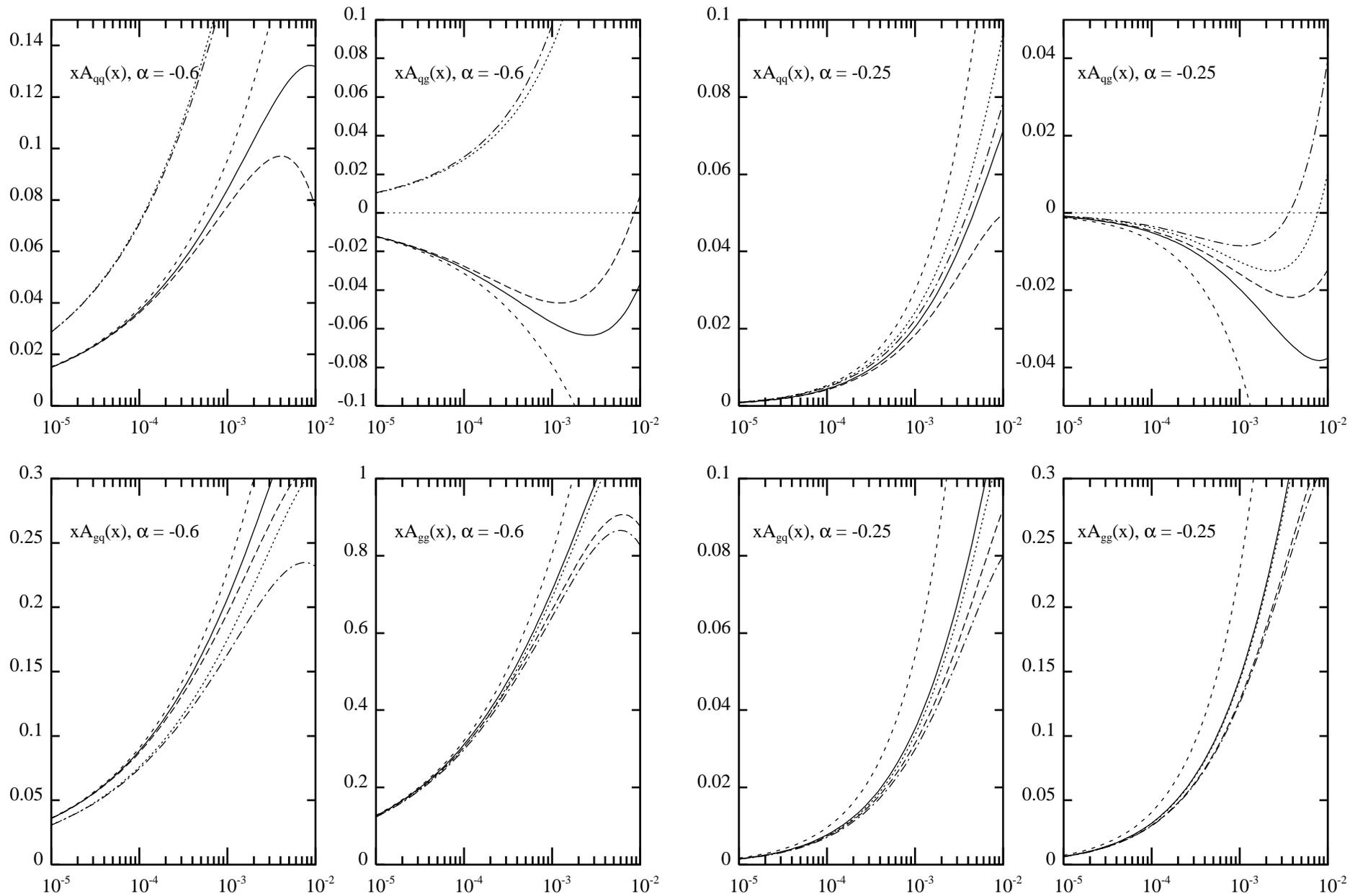
\item[Figure 2 ] Elements of the splitting matrix for the test
distribution $x^{\alpha}(1-x)^{\beta}$. Solid line: full splitting
functions for $\beta=4$, long-dashed line: same for $\beta=9$,
short-dashed line: most singular $x^{\alpha}$ contribution, dotted
line: leading-pole expanded splitting functions for $\beta=4$,
dot-dashed line: same for $\beta=9$. For better visibility, all
elements are multiplied by $x$.

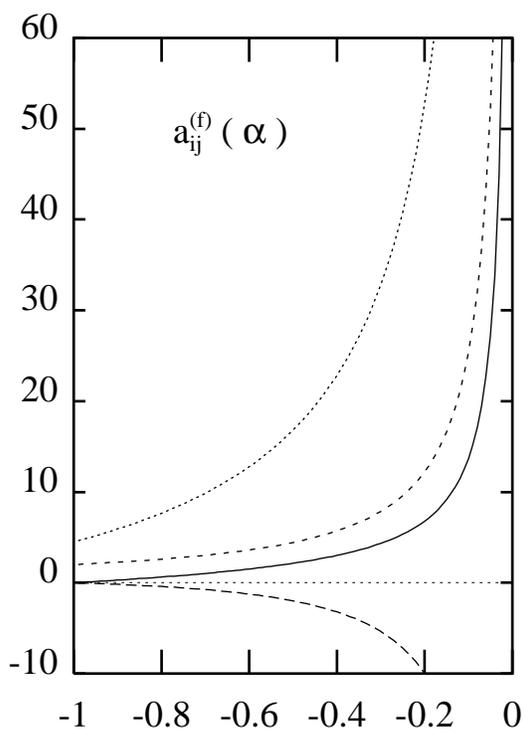
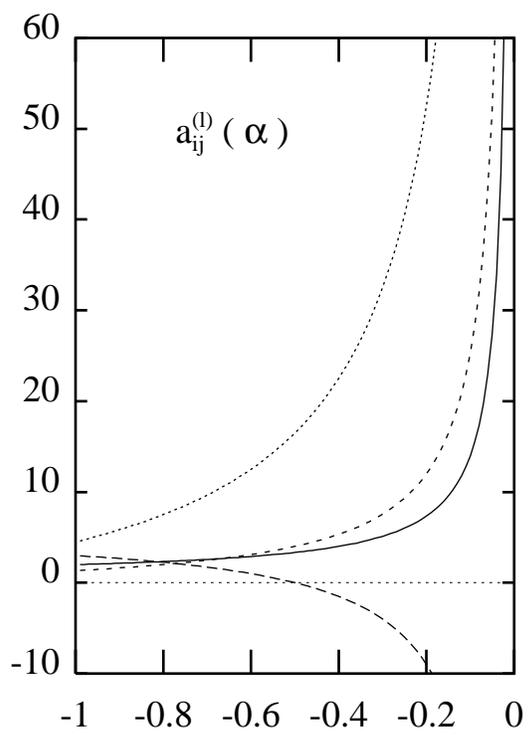
\item[Figure 3 ] Coefficients of the most singular pieces in the
splitting matrix for the full (left) and the leading-pole expanded
(right) splitting functions. Solid line: $a_{qq}$, long-dashed line:
$a_{qg}$, short-dashed line: $a_{gq}$, dotted line: $a_{gg}$.

\item[Figure 4 ] Evolution of test distributions
for gluons ($x\Delta G(x,Q^2)$),(a), singlet
quarks  ($x\Delta \Sigma (x,Q^2)$),(b) and valence quarks ($x\Delta
q_{val} (x,Q^2)$),(c)  as
described in the text.
Solid line: starting
distribution at $4 \; \mbox{GeV}^2$, long-dashed and dot-dashed line:
evolved distributions at $100 \; \mbox{GeV}^2$ for
different large-$x$ behaviour at $Q_0^2$, short dashed line: result of
$x^{\alpha}$ approximaton, dotted line: result of leading-pole
approximation.

\end{description}

\vfill
\newpage

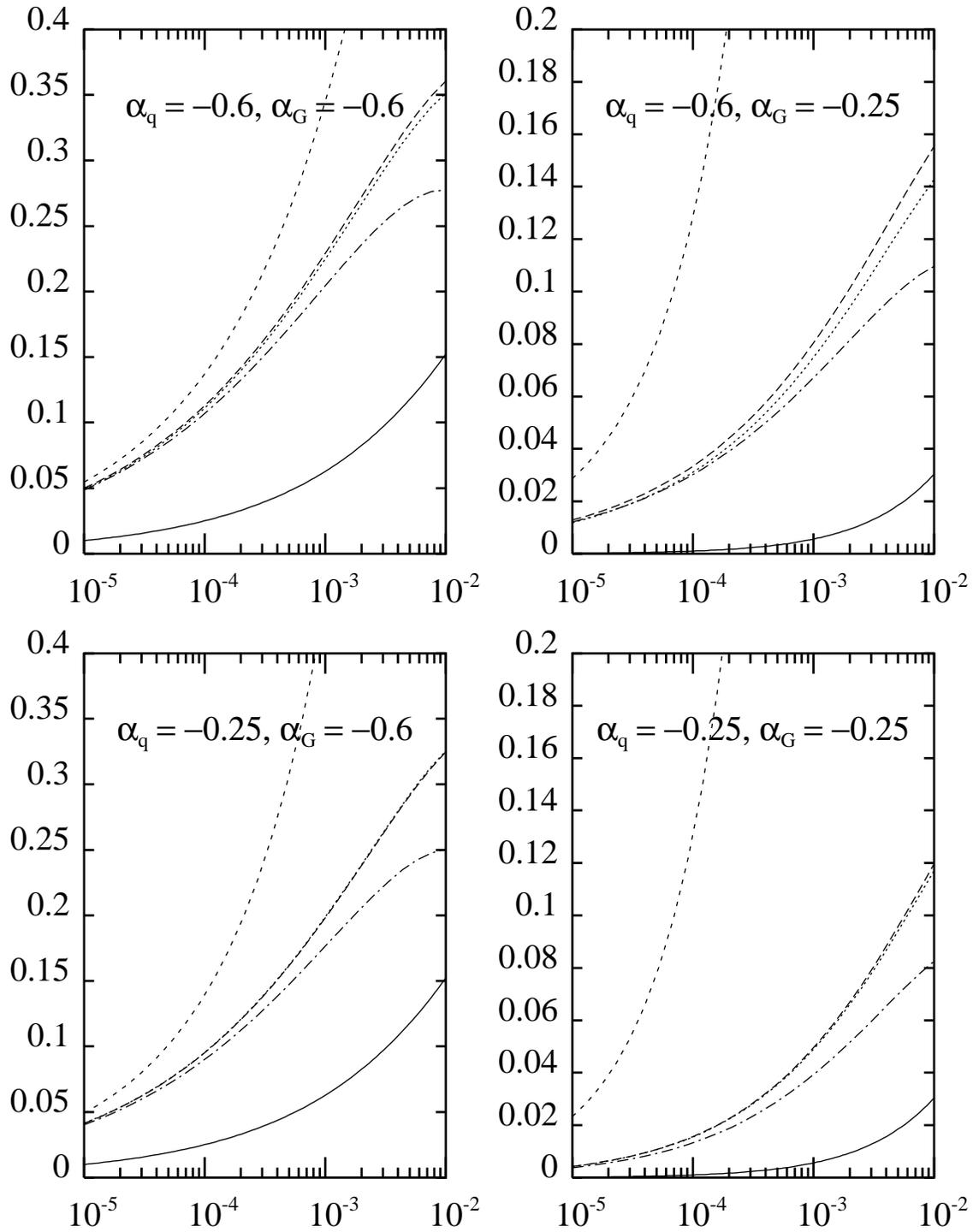
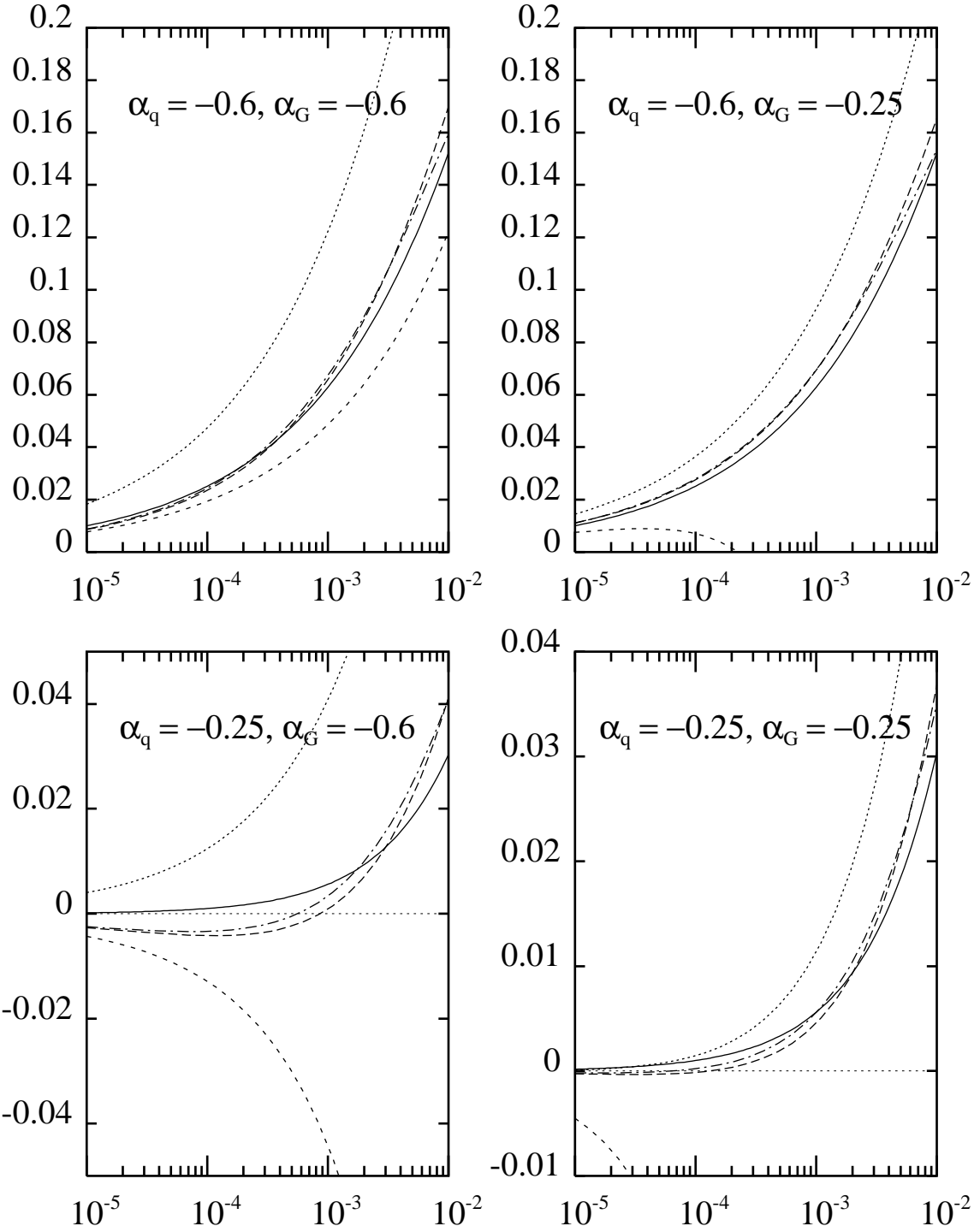
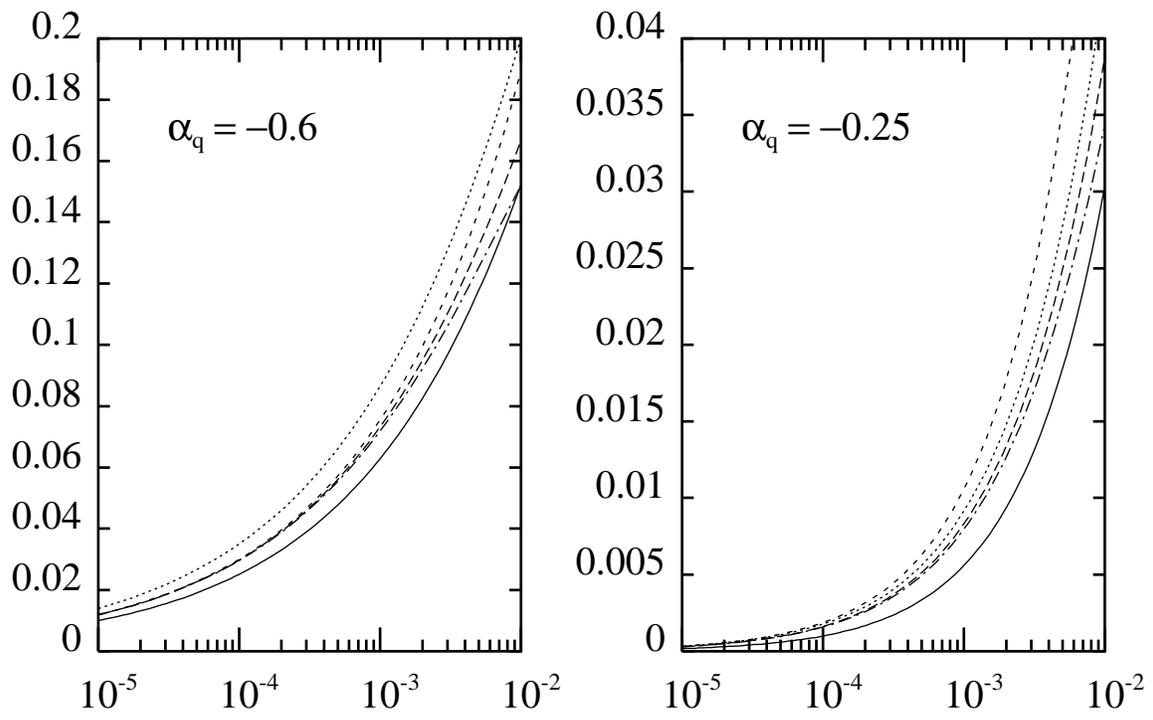
\end{document}